# Dose, exposure time, and resolution in Serial X-ray Crystallography.

**D. Starodub,[a*] P. Rez,[a] G. Hembree,[a] M. Howells,[b] D. Shapiro,[b] H. N. Chapman,[c] P. Fromme,[d] K. Schmidt,[a] U. Weierstall,[a] R. B. Doak,[a] J. C. H. Spence[a]**

[a]*Department of Physics, Arizona State University, P.O. Box 871504 Tempe, Arizona 85287-1504, USA,* [b]*Advanced Light Source, Lawrence Berkeley National Laboratory, Berkeley, California 94720, USA,* [c]*Lawrence Livermore National Laboratory, 7000 East Avenue, Livermore, California 94550, USA,* [d]*Department of Chemistry and Biochemistry, Arizona State University, P.O. Box 871604 Tempe, Arizona 85287-1604, USA. E-mail: dmitri.starodub@asu.edu*

**Abstract**

The resolution of X-ray diffraction microscopy is limited by the maximum dose that can be delivered prior to sample damage. In the proposed Serial Crystallography method, the damage problem is addressed by distributing the total dose over many identical hydrated macromolecules running continuously in a single-file train across a continuous X-ray beam, and resolution is then limited only by the available molecular and X-ray fluxes and molecular alignment. Orientation of the diffracting molecules is achieved by laser alignment. We evaluate the incident X-ray fluence (energy/area) required to obtain a given resolution from (1) an analytical model, giving the count rate at the maximum scattering angle for a model protein, (2) explicit simulation of diffraction patterns for a GroEL–GroES protein complex, and (3) the frequency cut off of the transfer function following iterative solution of the phase problem, and reconstruction of an electron density map in the projection approximation. These calculations include counting shot noise and multiple starts of the phasing algorithm. The results indicate counting time and the number of proteins needed within the beam at any instant for a given resolution and X-ray flux. We confirm an inverse fourth power dependence of exposure time on resolution, with important implications for all coherent X-ray imaging. We find that multiple single-file protein beams will be needed for sub-nanometer resolution on current third generation synchrotrons, but not on fourth generation designs, where reconstruction of secondary protein structure at a resolution of 7 Å should be possible with short (below 100 s) exposures.

## 1. Introduction

In order to solve the structure of proteins which are difficult to crystallize, we have proposed spraying them across a synchrotron X-ray beam and aligning them using the dipole moment induced by a near-infrared polarized laser (Spence and Doak, 2004; Starodub et al., 2005). All three orthogonal intersecting beams (the single-file protein beam, the alignment laser, and the X-ray beam) operate quasi-continuously (without synchronization) until adequate signal-to-noise ratio is achieved in the diffraction pattern, which is then read out. By rotating the polarization of an elliptically polarized laser, this process may then be repeated for many orientations to fill the 3D volume in reciprocal space with diffraction data. Other alignment methods, such as static electric or magnetic fields, or flow alignment have been considered and demonstrated (Bras et al., 1998; Koch et al., 1988), as well as employed in the field of birefringence measurements (Fredericq and Houssier, 1973). These alignment techniques may also be helpful to avoid the problem of orientation classification of diffraction patterns from single molecules in random orientations, which is the main difficulty arising for single molecule imaging using pulsed X-ray free-electron lasers(Chapman et al., 2006a; Huldt et al., 2003). The motion of the molecules does not affect the diffraction pattern if the illuminating wave field is approximately planar, so that if there is, for example, one molecule in the beam at any instant, the method is equivalent to diffraction from a single stationary molecule. The continuous replacement of this molecule by others, however, allows an arbitrarily long exposure time without radiation damage. For 20-μm diameter X-ray and laser beams, with a typical droplet beam velocity $v$ = 50 m/s, the transit time through the beam is $t$ = 400 ns. The radiation dose received by each protein during this time can be estimated by applying the Bragg's rule of weighted summation of monatomic photoabsorption cross sections for the elements composing a protein. Using tabulated data on photoabsorption cross sections (Henke et al., 1993), for a generic protein stoichiometry $H_{50}C_{30}N_9O_{10}S_1$ and density 1.35 gm/cm$^3$, that gives a mass absorption coefficient $\mu$ of 9.6 cm$^2$/g at the X-ray energy $E$ = 8 keV, required to obtain a near- atomic resolution. The dose which sets the radiation damage limit at atomic resolution is $D_L$ = 2×10$^7$ Gy (Henderson, 1995). That assumes that an ejected photoelectron passes through surrounding bulk material causing damage, and therefore gives a lower limit on acceptable dose for the isolated biomolecules in Serial Crystallography, where the photoelectrons deposit only a small fraction of their energy before escaping into vacuum. Then the lower limit of the maximum tolerable flux for 8-keV X-rays is $I_0 = D_L/\mu E t = 4.1 \times 10^{10}$ photons s$^{-1}$ nm$^{-2}$. Although at a lower X-ray energy the absorption coefficient increases, the radiation damage dose at the increased scale of resolution, feasible at this energy, increases as well. Therefore, the maximum tolerable flux does not increase dramatically as X-ray energy decreases. This beam flux is far beyond the capabilities of any existing or projected X-ray sources. Therefore, the resolution, achievable in Serial Crystallography, is not limited by radiation damage, and depends chiefly on the effectiveness of alignment process (Spence et al., 2005) and time available for data collection.

If there is no interference between X-rays scattered from different molecules, then the scattered intensity from a single-file train of macromolecules with separation $L$ traveling across an X-ray beam of diameter $D_B$ = 20 μm is proportional to the number of molecules falling within the beam at any instant $M = D_B/L$. We further assume that all $M$ molecules are perfectly aligned. For a monodispersed Rayleigh droplet beam, the droplet diameter is about twice that of the column of liquid from which they form by a necking instability (Rayleigh, 1878), and the spacing between droplet centers is about twice their diameter. Therefore, the 1-μm liquid column produced by a Rayleigh droplet source gives $L$= 4 μm and $M$ = 5, resulting in an 80% reduction in exposure time over single molecule exposure at the same resolution. In order to increase the scattering intensity, the design of "shower-head" aerodynamically formed multiple-jet nozzles is also under active development (Weierstall et al., 2007). Experiments are planned with an average of one protein per droplet, and also with many proteins per droplet. Data will also be collected using an average of one sub-micrometer protein crystallite in each droplet. Even without alignment, the resulting "powder protein data" might be solved by molecular replacement methods using the iterative flipping algorithm (Wu et al., 2006). In this paper we treat mainly the case of one molecule per droplet, and assume that all water except a few-monolayer jacket of vitreous ice has been removed, as in recent research on proteins using electrospray spectroscopy (Sobott et al., 2005), so that the ice-jacket effects can be ignored. Inclusion of the ice background may increase the required dose by almost one order of magnitude, however by choice of flight distance the jacket thickness may be reduced to zero.

The purpose of this paper is to provide realistic estimation of exposure time required for diffractive imaging of biological macromolecules. We perform simulations of the diffraction patterns for a sample object at various exposure times, and then apply the iterative procedure to solve the phase problem for charge density reconstruction in order to determine the relationship between exposure time $\Delta t$ and resolution $d$ in reconstructed image. The results are compared with power law estimates derived from simple scattering models.

Our project grew out of earlier work on coherent diffractive imaging (Marchesini et al., 2003a) based on a soft X-ray undulator beam [beamline 9.0.1 at the Advanced Light Source (ALS)], using a zone-plate as a monochromator. Diffraction from virus particles was intended, requiring 500 nm spatial coherence and the high flux, made possible only by an undulator operating in the soft X-ray region. For large proteins or macromolecular assemblies at 20 Å resolution, shorter wavelengths and less coherence are needed, so that our simulations here are given for the new COSMIC beamline at the ALS with an undulator optimized for producing soft X-rays in the energy range 0.25 – 3 keV, for a new coherent 2-6 keV undulator beamline at the Advanced Photon Source (APS), and for the energy-recovery linac (ERL) source planned at Cornell.

**2. Relationship between resolution and exposure time.**

Related treatments of the relationship between exposure, dose, resolution and beam energy for X-ray microscopy have been given previously (Howells et al., 2005; Marchesini et al., 2003a; Shen et al., 2004). That work is based on calculation of the imaging dose (energy absorbed per unit mass) required to collect statistically significant data at a given resolution. If this dose is smaller than that known to destroy structural detail of a given size, this resolution is considered feasible. Otherwise, the resolution limit is determined by the dose that destroys detail of a given size. A statistically reliable photon count $P$, chosen for this dose, may be found in either of two ways. The first approach is to calculate the total number of photons scattered into the detector from a single sample voxel with linear dimensions $d/2$ where d is the resolution.. (These counts will subsequently be phased and recombined computationally into one resolution voxel in the real space reconstruction or image). Alternatively, one may calculate the number of photons scattered by the entire object of size $D$ into one detector pixel at a scattering angle corresponding to the resolution of interest. The first method is independent of molecular size, the second is not. Both methods depend on the structure of the object (in the first method, the result depends on which voxel is chosen), so that resolution is here a property of the sample as well as the instrument. In the first approach (Howells et al., 2005), one can simply integrate the signal, scattered by a spherical voxel of diameter $d/2$, to get (Kirz et al., 1995)

$$P = \frac{\pi}{128} r_e^2 \lambda^2 d^4 |\rho|^2 I_0 \Delta t \qquad (1)$$

in the limit $\lambda \ll d$, where $r_e = 2.82 \times 10^{-6}$ nm is the classical radius of electron, $\rho = n_a (f_1 + i f_2)$ is the effective complex electron density of a matter with atomic concentration $n_a$ and complex atomic scattering amplitude $f_1 + i f_2$, $\lambda$ is the X-ray wavelength, $I_0$ is the incident X-ray flux, and $\Delta t$ is the data acquisition time. Then the dose, proportional to the incident X-ray beam fluence, scales with resolution as $d^{-4}$. The required exposure does not depend on detector size. In the second approach, an *incoherent* sum over the object volume of the scattered intensities from the resolution-elements (voxels) of size $d$ into a detector pixel corresponding to resolution $d$, is (Shen et al., 2004)

$$P = \frac{3}{4\pi^3} r_e^2 \lambda^2 d^3 D |\rho|^2 I_0 \Delta t \qquad (2)$$

(By contrast, we assume a coherent sum below). This result depends on the object size, and the shape of the resolution element. Note that if the latter were cubic, the scattered intensity at scattering vector $q = 2\pi/d$ corresponding to the resolution limit would be zero.

The generally accepted requirement for a statistically reliable measurement of signal $P$ is that that signal exceeds the background noise level by a factor of five (Rose, 1948). Since the input to the numerical phase retrieval algorithms involves the modulus of the scattered amplitude rather than intensity, for Poisson noise this implies $\sqrt{P}/(\Delta\sqrt{P}) = 2P/\Delta P = 2\sqrt{P} = 5$, or $P = 6.25$. Successful 3D reconstruction from experimental diffraction patterns has been reported at the photon count of just 1 photon/pixel at the highest achieved resolution (Chapman et al., 2006b). To be consistent with previous work (Shen et al., 2004), we choose $P = 5$ for further discussion.

Because the coherence patch of the synchrotron is larger than our biomolecule, we assume that the statistical accuracy of a diffraction pattern is defined by the *coherent* scattering from the entire object at the angle that corresponds to the required resolution. For convenience we start by considering the scattering from a single organic spherical object of radius $R = D/2$. The incident X-ray wave vector is **k**, the scattered wave vector **k'** and the scattering vector **q**. The vector **q** connects the (000) point with other points on the momentum and energy-conserving Ewald sphere of radius $k$:

$$\mathbf{q} = \mathbf{k'} - \mathbf{k},$$

with the maximum value $q_{max}$ defined by the maximum scattering angle allowed by the detector geometry. To obtain a full 3D reconstruction, diffraction patterns from all object orientations must be recorded, in order to fill a sphere of radius $q_{max}$ in reciprocal space. These intensities measured on the Ewald sphere can then be redistributed onto a regular Cartesian grid by interpolation. As our object is coherently illuminated by X-rays, with incident electric field $E_0$, the electric field amplitude at a distance $r$ in a direction specified by **q** is

$$E'(\mathbf{q}) = \frac{r_e}{r} E_0 \sin\psi \int \rho(\mathbf{r}) \exp(i\mathbf{q}\cdot\mathbf{r}) d\mathbf{r} \qquad (3)$$

where $\rho(\mathbf{r})$ is the charge density and $\psi$ represents the angle between the electric field and the scattered direction (a polarization term). If the sphere had uniform charge density $\rho$, then the Fourier transform in Eq. (3) could be evaluated as

$$f_x(q) = 4\pi\rho R^3 \frac{[\sin(qR) - (qR)\cos(qR)]}{(qR)^3}. \qquad (4)$$

In terms of intensities and a cross section the scattering equation can be written as

$$I(q) = r_e^2 |f_x(q)|^2 \sin^2\psi \Delta\Omega I_0 \qquad (5)$$

To solve the phase problem using the scattered intensity pattern based on the iterative Fienup (1982) algorithm, the object must be embedded in a known matrix of extent $sD$, with sampling ratio $s = 2^{1/3}$ for 3D reconstruction and $s = 2^{1/2}$ for 2D reconstruction. Then the pixel size in reciprocal space is $\Delta q = \pi/sR$. If we consider scattering by relatively small angles, then the solid angle subtended by a pixel is

$$\Delta\Omega = \left(\frac{\Delta q}{k}\right)^2 = \left(\frac{\lambda}{2sR}\right)^2. \qquad (6)$$

As seen from Eq. (4), the shape function for a uniform charge density falls as $q^3$. Additionally, the atomic scattering factor also decreases for a larger scattering vector. This means that reconstructing an object to a given resolution $d$ requires that there be statistically significant counts in a pixel at $q_{max} = 2\pi/d$. If the detector has $N \times N$ pixels and its center is on the axis of the incident beam, at the edge of the detector $q_{max} = N\Delta q/2$. Then the expression for resolution becomes

$$d = \frac{4sR}{N} \qquad (7)$$

From Eqs. (4)–(6), we get for the scattered photon count $P$ at the pixel corresponding to the scattering vector $q$

$$P = I(q)\Delta t = \frac{4\pi^2 r_e^2 \lambda^2 |\rho|^2 I_0 \Delta t}{s^2 R^2}\left(\frac{\sin(qR)-(qR)\cos(qR)}{q^3}\right)^2 \qquad (8)$$

The oscillation period of the term in parentheses slightly exceeds the pixel size, and becomes equal to that at $s = 1$. Averaging of this term in radial direction over the pixel size gives for small $s$

$$\frac{sR}{\pi}\int_{q-\pi/2Rs}^{q+\pi/2Rs}\left[\frac{\sin(qR)-(qR)\cos(qR)}{q^3}\right]^2 dq \approx \frac{1}{2}\frac{R^2}{q^4}. \qquad (9)$$

Combining Eqs. (8)–(9), we obtain for the number of counts in time $\Delta t$ at the pixel corresponding to resolution $d$:

$$P = \frac{1}{8\pi^2 s^2} r_e^2 \lambda^2 d^4 |\rho|^2 I_0 \Delta t. \qquad (10)$$

This expression has the same functional dependence as that obtained by Howells et al (Howells et al., 2005), but with a different numerical prefactor. We note in particular the power-law scaling with $d$ and $\lambda$. A similar result is obtained in the phase-grating approximation, applied to one voxel. Here the phase shift $\theta = r_e \rho \lambda d$, which produces a cross-section $d^2\theta^2$, as above.

**3. Scattering simulation.**

A more detailed analysis, extending to the important sub-nanometer resolution range, is possible using a direct calculation of the X-ray diffraction pattern based on atomic X-ray scattering factors. This allows the effects to be considered of three-dimensional atomic structure, detector size, noise, and stability of our iterative algorithm for solution of the phase problem. As the test object for our simulations we choose the asymmetric *E. coli* chaperonin GroEL$_{14}$–GroES$_7$–(ADP·AlF$_x$)$_7$ protein complex, constituted of 59,276 atoms. GroEL contains 14 identical subunits of molecular mass 58 kDa, and GroES contains 7 subunits of molecular mass 10 kDa. They form a structure consisting of three distinctive rings. The length of the complex is 20 nm, and diameter 14.5 nm. The 3D structure of the complex at 2.8 Å resolution has been reconstructed by X-ray crystallography (Chaudhry et al., 2004) and was obtained from the Protein Data Bank (entry 1SVT). A sketch of the scattering geometry is shown in Fig. 1. The detector is a two-dimensional 512×512 array of equidistant pixels of linear size $a$, located at a distance $l_{00}$ from the sample, which limits the scattering angle at its edges to a resolution of a few ångströms. The position of a pixel with indices $i$ and $j$ relative to the sample is defined by the zenith angle $\theta_{ij}$ and azimuth angle $\varphi_{ij}$, which also determine the distance $l_{ij}$ between this pixel and the sample. Then the diffraction pattern is formed by the polar gnomonic projection of the points lying on Ewald sphere onto the flat detector screen. According to Eq. (5), for an incident plane wave of intensity $I_0$ with wavelength $\lambda$ the scattered photon count per unit time in the given pixel is given in the single-scattering (Born) approximation as

$$I_{ij} = r_e^2 A(q_{ij}) A^*(q_{ij}) \sin^2(\psi_{ij}) \Delta\Omega_{ij} I_0,$$

(11)

with a sample scattering amplitude

$$A(q_{ij}) = \sum_k f_k(q_{ij}) \exp(i\mathbf{q}_{ij}\mathbf{r}_k),$$

(12)

where $q_{ij} = 4\pi\sin(\theta_{ij}/2)/\lambda$ is the scattering vector corresponding to detector pixel ($i,j$), which subtends solid angle $\Delta\Omega_{ij} = a^2 \cos(\theta_{ij})/l_{ij}^2$ at the angle $\theta_{ij}$. $\mathbf{r}_k$ is the position vector of the $k^{\text{th}}$ atom in the sample, and the summation is performed over all the atoms of the sample. For the undulator odd harmonics the X-ray beam is linearly polarized and polarization term is

$$\sin^2(\psi_{ij}) = 1 - \sin^2(\theta_{ij})\cos^2(\varphi_{ij}).$$

The scattering amplitude for the $k^{\text{th}}$ atom is (Henke et al., 1993)

$$f_k(q) = f_k' + if_k'' - \Delta f_k(q),$$

where the last term describes the angular dependence of the atomic form factor:

$$\Delta f_k(q) = Z - \tilde{f}_k(q),$$

and $\tilde{f}_k(q)$ is the empirical approximation of tabulated data by four-Gaussian fitting (Doyle and Turner, 1968), satisfying condition $\tilde{f}_k(0) = Z$. We note that, if absorption is neglected so that $f_k(q)$ is real, then $A(-\mathbf{q}) = A^*(\mathbf{q})$, and the charge density obtained by Fourier transform of the sample scattering amplitude is real.

To satisfy the Shannon sampling requirement $\Delta q = 2\pi/sD$, the maximum allowed beam angular spread at the sample should be equal to $\theta_c = \lambda/2sD$ (Spence et al., 2004). The requirement that the photons with bandwidth $\Delta\lambda$, scattered at the same scattering vector, diverge by no more than half a detector pixel, results in the condition $\Delta\lambda/\lambda < 2/N$ (Chapman et al., 2006b; Spence et al., 2004), where $N$ is the number of pixels along one Cartesian axis. According to the Eq. (7) with the sampling ratio of $s = 2^{1/2}$, 7 Å resolution requires 81 pixel. Then the desired energy bandwidth is 2.5%, significantly larger than the available energy spread. The requirements for the beam angular spread and a spot size of A = 20 μm×20 μm define the volume of the beam transverse phase space acceptable for the scattering experiment. Since the undulator cannot fill this space, we assume that all phase space of the actual X-ray beam can be used.

We perform simulations of the diffraction patterns from the GroEL–GroES protein complex using the parameters of three new X-ray beamlines that will become available in the near future. The first one is the recently proposed COSMIC beamline at Section 7 of ALS, which will provide brightness $B$ larger than $10^7$ photons s$^{-1}$ nm$^{-2}$ mr$^{-2}$/0.1%BW in the energy range of 0.25–3 keV. For the best resolution it will be adventurous to operate at the maximum possible energy. The width and angular spread of the X-ray beam are determined by its convolution with the electron beam in the storage ring (Spence and Howells, 2002). Using parameters for the COSMIC undulator at 3 keV, we find the X-ray root mean square (rms) horizontal width of $\sigma_{Tx}$ = 0.293 mm and vertical width $\sigma_{Ty}$ = 8.53 μm, rms angular spread of $\sigma_{Tx'}$ = 25.9 μr and $\sigma_{Ty'}$ = 14.9 μr in horizontal and vertical direction, respectively. Then the X-ray intensity at the sample is $I_0 = (2\pi)^2 B \sigma_{Tx} \sigma_{Ty} \sigma_{Tx'} \sigma_{Ty'}/A = 0.95 \times 10^6$ photons s$^{-1}$ nm$^{-2}$ at 0.1% energy bandwidth. Taking into account the estimated 76% loss in the optical system and adjusting for the maximum bandwidth of $\Delta\lambda/\lambda = 1/151$ possible at 3 keV, we finally get $I_0 = 1.5 \times 10^6$ photons s$^{-1}$ nm$^{-2}$. Since the phase space volume in one dimension of a single mode beam is $\lambda/4\pi$, the total number of modes in the beam is 231×3.87 = 894.

The second example considers the projected undulator source at APS, which will operate in the energy range between 2 and 6 keV. A recent measurement of the similar undulator beam at Sector 7 of APS, focused into a ten micrometer spot, gave 6×10$^{12}$ incident photons s$^{-1}$ with beam divergence of 1.4 mrad at 0.01% energy bandwidth at 14.3 keV (Young et al., 2006). The same analysis as for the COSMIC beamline gives the optimized beam intensity of 1.8×10$^6$ photons s$^{-1}$ nm$^{-2}$ at 5.4 keV. For the harder X-rays, suitable for near atomic resolution, we use a flux of 3×10$^8$ photons s$^{-1}$ nm$^{-2}$ at 8 keV, corresponding to the proposed ERL beamline at Cornell university (Shen et al., 2004).

The secondary structure of proteins (α-helices) can be resolved at resolution of $d$ = 7 Å, which sets the lower limit for the largest measured scattering vector as 0.9 Å$^{-1}$. The sampling ratio can be found from Eq. (7) as $s = Nd/2D$, and oversampling (relative to the minimum acceptable sampling ratio) for 2D projection is then $Nd/2^{3/2}D$. Thus, for a detector whose linear size is $N$ = 512 pixels and $D$ = 200 Å the diffraction pattern is oversampled by a factor of 6.3. A simulated diffraction pattern, on the 512×512 grid for one molecule in the 8-keV X-ray beam, is presented in the left panel of Fig. 2. The right panel shows the scattered intensity per pixel, averaged over the azimuth angle, as the function of scattering vector for incident beam energies of 3.0, 5.4 and 8.0 keV with intensities of 1.5×10$^6$, 1.8×10$^6$ and 3×10$^8$ photons s$^{-1}$ nm$^{-2}$, respectively. If normalized to the same incident flux, the ratio of the integrated scattered intensities (scattering cross sections) for 3.0, 5.4 and 8.0 keV is 7.3:2.2:1. This is close to the ratio 7.1:2.2:1, predicted by a $\lambda^2$ scaling of scattered coherent flux with X-ray wavelength, according to Eq. (1). Lower X-ray energy can require a closer distance between detector and sample in order to collect the data at large scattering angles. In this case the pixel size in reciprocal space near the center of the detector will be larger than the average, and failure to interpolate the diffraction pattern onto a regular grid of scattering wave vectors would result in a "stretched" reconstructed object. Additionally at large scattering angles the count rate is affected due to reduction of the solid angle subtended by a pixel near the edge of the detector by a factor of cos ($\theta$). The dashed horizontal line in Fig. 2(b) corresponds to 5 counts per pixel after exposure of 100 s at minimum required sampling ($N$ = 81) for one sample in the X-ray beam at any instant. Therefore, its intersections with the scattering curves at different X-ray energies determine the resolution achievable under these conditions. In particular, the full reconstruction of the secondary protein structure can be predicted at the future ERL source, based on the requirement of 5 counts in a pixel at the maximum scattering angle, while the low resolution envelope could be obtained at the APS and ALS (with the resolution of 27 Å and 18 Å, respectively). For $M$ proteins in the beam at any instant, the count rate is multiplied by $M$, since no interference occurs between different molecules. Therefore, with a reasonable assumption of $M$ = 15-30, a resolution of 7 Å is feasible at the APS and ALS as well. Object reconstruction from these diffraction patterns (with noise added), as described in the next section, shows that larger exposure times than predicted here are actually required for the intended resolution. Note that full 3D reconstruction requires that the collected data be assigned to points on the Ewald sphere, which is swept through reciprocal space (by rotating the sample) to fill a 3D volume. Using diffraction patterns from different protein orientations independently would then increase tremendously the time required for data acquisition. However, if the correlation between various projections is taken into account for 3D reconstruction according to the dose fractionation theorem (Hegerl and Hoppe, 1976), the dose required for each projection in the 3D imaging will be reduced. The Hegerl – Hoppe theorem states that the full 3D reconstruction of an object requires the same total dose (distributed over many orientations) as the reconstruction of a single 2D projection at the same resolution. It is important to note that the scattering signal must be statistically reliable to resolve a single 3D pixel in the 2D projection, rather than a 2D pixel formed by summation of 3D pixels along the projector line. (McEwen et al., 1995). Thus the dose, required for resolution $d$, can be determined by considering scattering from an object slice of thickness $d/2$. The 3.5 Å thick slice of the GroEL contains about 1/40 of all atoms in the complex. Then the average count rate for that slice at the scattering angle, corresponding to resolution $d$, is 1/40 of that calculated above for the whole object. Correspondingly, the total dose required for the GroEL 3D reconstruction is 40 times larger than the dose needed for the statistically accurate measurement of one orientation.

In the inset of Fig. 2(b) the scattering curves for 8 keV photons with an incident flux of $3 \times 10^8$ photons s$^{-1}$ nm$^{-2}$ (ERL) in the planes of $q_x = 0$ (solid line) and $q_y = 0$ (dash line) are re-plotted using log-log coordinates. They are extracted from the 256×256 grid (sampling ratio $s = 4.48$). As expected for an asymmetric object, at smaller scattering vectors the scattering curves are feature-rich and highly inhomogeneous. In particular, a pronounced peak at $q_y = 0.15$ Å$^{-1}$ corresponds to the ring structure of the GroEL complex in $y$ direction with a period of about 40 Å. It should be clear that in this scattering vector region it would be difficult to rely upon the general form of a power law (as derived in the previous section) for the required flux estimation, since the scattering curve in this region cannot be fitted by a power law. Due to the sharp peak at 0.15 Å$^{-1}$, this feature will dominate even in very noisy diffraction patterns, giving rise to a disk-like structure. The scattering curves become relatively featureless and independent of azimuth angle only at the highest scattering vectors, and then they can be approximated using a power law, resulting in the power scaling of the required exposure time with resolution, described in the previous section.

In the discussion above, we have defined resolution by the highest scattering angle at which statistically accurate data above background can be found in a detector pixel. This treatment does not take into account the stability of the reconstruction algorithm, used for phase retrieval, with respect to statistical fluctuations. Therefore, it gives a lower limit for the required exposure. In the next section we apply the Hybrid Input Output (HIO) algorithm to reconstruct the high-resolution structure of the GroEL complex, and quantitatively investigate resolution as a function of incident fluence.

**4. Coherent transfer function for HIO reconstruction.**

Because the HIO algorithm is known to be more effective for real-valued objects, where a strong positivity constraint can be applied, we limit our consideration to a real object. In the general case the Fourier transform of the scattering amplitude $A(\mathbf{q})$, collected on a 2D grid, would not be real because one measurement cuts reciprocal space along the curved Ewald sphere, which does not contain points with inverted coordinates, and therefore the condition for object reality $A(-\mathbf{q}) = A^*(\mathbf{q})$, while satisfied by a tomographic data set collected in three dimensions, is not met on a 2D grid. For simplicity, we do not consider such a full 3D reconstruction, and to be consistent with the requirement of object reality, we use the diffraction pattern from a 2D projection of the GroEL-GroES electronic density, calculated by setting coordinate $z = 0$ in Eq. (12). This also avoids the de-focusing effects in the projection approximation for 3D objects due to the curvature of the Ewald sphere. We assume that the atomic scattering amplitude is equal to the number of electrons in atom $Z$, thus neglecting absorption and any angular dependence of scattering amplitude, which is justified for high energy photons (here 8 keV) and scattering at small angles. The projection of the object electronic density is given in the inset to Fig. 3 (left panel). The scaling bar length corresponds to 35 Å. The diffraction pattern was calculated on a 256×256 grid with a maximum wave vector transfer of $q_{max} = 0.9$ Å$^{-1}$ (sampling ratio $s = 4.48$). We found that application of the HIO algorithm to a 128×128 grid ($s = 2.23$) results in a smaller percentage of successful reconstructions.

The HIO iterative algorithm (Fienup, 1982) with reality and positivity constraints is described by a recursion relationship

$$g_{n+1}(\mathbf{r}) = \begin{cases} \hat{P}_M g_n(\mathbf{r}) & \text{if } \mathbf{r} \in S \wedge \Re(\hat{P}_M g_n(\mathbf{r})) \geq 0 \\ (1 - \beta \hat{P}_M) g_n(\mathbf{r}) & \text{otherwise} \end{cases},$$

(13)

where $g_n(\mathbf{r})$ is the reconstructed object in real space after $n$-th iteration, $S$ defines a support such that $g(\mathbf{r} \notin S) = 0$, and the feedback parameter is $\beta = 0.9$. The projector operator $\hat{P}_M$ determines the projection of the Fourier transform of the reconstructed object on the reciprocal space subset satisfying the modulus constrain defined by the measured scattered intensities:

$$\hat{P}_M g_n = F^{-1}\left(\frac{F(g_n)}{|F(g_n)|}\sqrt{I(\mathbf{q})}\right).$$

(14)

Here $F(g_n)$ denotes the operation of Fourier transform. We use 1000 cycles of the HIO iterations followed by 5 cycles of the error-reduction (ER) algorithm $g_{n+1}(\mathbf{r}) = \hat{P}_S \hat{P}_M g_n(\mathbf{r})$, where the support projector is:

$$\hat{P}_S g_n(\mathbf{r}) = \begin{cases} g_n(\mathbf{r}) & \text{if } \mathbf{r} \in S \wedge \Re(g_n(\mathbf{r})) \geq 0 \\ 0 & \text{otherwise} \end{cases}.$$

(15)

The quantitative measure of the iterative process convergence is the error metric in real space (equal to the normalized amount of charge-density remaining outside the support)

$$E_n^S = \sqrt{\frac{\sum_{\mathbf{r} \notin S}|g_n(\mathbf{r})|^2}{\sum_{\mathbf{r} \in S}|g_n(\mathbf{r})|^2}},$$

(16)

and in reciprocal space

$$E_n^M = \sqrt{\frac{\sum_\mathbf{q} \left\| F(g_n) \right| - \sqrt{I(\mathbf{q})} \right|^2}{\sum_\mathbf{q} I(\mathbf{q})}} .$$

(17)

We determined the support by convolution of the known object, used for the calculation of the diffraction pattern, with a Gaussian whose half width at half maximum was 3 pixels (10.5 Å), and a subsequent cut off at 5% of the maximum object charge density. The area outside the support is marked by the gray color on the right panel of the inset to Fig. 3. Because the support is relatively loose, the actual sampling ratio $s$ (which should be more correctly defined relative to the support size rather than the object size, as we do here) is somewhat smaller than indicated. Even without using the support, the HIO algorithm provides the low-resolution sample structure and external boundary. This implies that the Shrinkwrap algorithm (Marchesini et al., 2003b) could be applied if the support were unknown. That algorithm dynamically refines an initially loose support using intermediate reconstructions after a series of iterative steps.

The first object estimation was determined by applying the support projector given by Eq. (15) to the Fourier transform of the measured modulus of the scattered amplitude $\sqrt{I(\mathbf{q})}$ with random phases $\varphi(\mathbf{q})$. To ensure the reality of this Fourier transform, the condition $\varphi(-\mathbf{q}) = -\varphi(\mathbf{q})$ was enforced. In spite of the support asymmetry, sometimes the reconstructed image appeared in the inverse orientation. Though usually it rotates to the correct position after a sufficiently large number of iterations, in order to facilitate the convergence rate the first 100 iterations are performed additionally using the same set of random phases, but with reversed signs. Then the reconstructed object with the larger error, which has a wrong orientation, is rejected, and the rest of the iterations are done using the remaining object with the correct orientation.

Fig. 3 shows the behavior of the rms error, defined by Eq. (16), in a single reconstruction procedure for three reconstructions with different initial phases. In all cases, after a few iterations the error drops to $E^S \leq 0.1$. The successful reconstructions (solid lines, rate of success is about 85%) are characterized by a step-like decrease of the error by about a factor of 2 at some point (in Fig. 3, after around 400 and 800 iterations), which is accompanied by decreasing of the error standard deviation. Reconstructions that do not converge to the correct solution have a persistently high and noisy error (open circles). Before averaging over successful reconstructions, the images must be re-aligned to accommodate for the origin ambiguity produced by different random starting phases. This was done in two ways: by adjusting the image position in real space and the phases in reciprocal space. The reconstructed image with the smallest rms error was chosen as a reference. Then in real space, each remaining image was translated to the position where its cross-correlation with the reference image has a maximum, in order to minimize the rms error between this and the reference images (Fienup, 1997). In reciprocal space, the linear shift of the reconstructed object is given by the slope of the difference map $\Delta\varphi(\mathbf{q})$ between the diffraction amplitude phases of this and the reference images. We compute the slopes $\Delta\varphi_x = \partial\Delta\varphi/\partial q_x$ and $\Delta\varphi_y = \partial\Delta\varphi/\partial q_y$ using the least square linear fit in the $q_x$ and $q_y$ directions of the central part of the phase difference map, where the noise of the recovered phases has a lowest value. In most cases, we define the central data segments for linear fit by the condition that the correlation coefficient for these segments, reflecting their linearity, is set equal to 0.9. Then the image translation along *x*-axis in real space (in pixels) is determined as $X = \Delta\varphi_x N/2\pi$, and similarly for the *y* direction. The image averaged over 171 successful reconstructions (out of 200), adjusted using the cross-correlation function, is shown in the right inset to Fig. 3. It clearly repeats the original image structure. In particular, the details of the top *trans* GroEL ring, medium *cis* GroEL ring, and bottom GroES cap can be observed.

To test the stability of the HIO algorithm convergence with respect to the noise level, we introduced shot noise for the number of photons collected by a detector, described by a Poisson distribution of counts in each pixel:

$$f_{ij}(k) = \frac{\exp(-s_{ij})s_{ij}^k}{k!},$$

where *k* is the integer number of counts in the (*i*, *j*) pixel, and $s_{ij} = I(q_{ij})M\Delta t$ is the expected number of counts in this pixel after exposure time $\Delta t$, determined from the calculated diffraction pattern. We assume an incident photon flux of $3\times 10^8$ photons s$^{-1}$ nm$^{-2}$ (ERL) and one sample in the beam at a time $M = 1$. Then the phase retrieval algorithm was applied as described above. For each exposure time, 200 independent reconstructions have been run, and 15% of the reconstructions with the highest error have been rejected. The real space rms error after the final iteration step, calculated according to Eq. (16) and averaged over successful reconstructions, is shown in Fig. 4 (a) as a function of exposure time. The mean error steadily increases as the input diffraction patterns become noisier, roughly following the power dependence on the counting time with the exponent of -0.28, as indicated by the fitting line. The images have been adjusted by either their positions or phases, as described above, and then averaged. The result is shown in Fig. 5 for both methods of image adjustment. Using the cross-correlation function in real space gives somewhat better images at low exposures. The details of the ring structure remain consistent at exposures as low as 10 s, but eventually they become completely smeared out at an exposure of 1 s, which is attributed to the fast growth of the HIO process instability at this counting time. The failure of the reconstruction algorithm is also reflected in the behavior of the error distribution, shown in Fig. 4 (b). In a large range of the longer exposure times, the errors of independent reconstructions have a very narrow and asymmetric distribution, which suddenly broadens as exposure decreases from 10 s to 1 s, indicating stagnation of the algorithm.

Visual examination of Fig. 5 allows one to follow the change of resolution in response to exposure time. Quantitative measure of resolution can be provided by the analysis of transfer function (TF). If the Fourier transform of the object $G(\mathbf{q})$ is considered as the output of the phase retrieval algorithm, then its TF for diffraction amplitude modulus can be defined as the ratio of the modulus of the output averaged over independent reconstructions to the modulus of the ideal scattered amplitude (Shapiro et al., 2005):

$$TF(q) = \frac{\left\langle \left|\left\langle G(\mathbf{q})\right\rangle\right|\right\rangle_\varphi}{\left\langle \sqrt{I(\mathbf{q})}\right\rangle_\varphi}$$

(18a)

Here the Fourier transform moduli have been averaged over azimuth angle prior to the TF

calculation. Averaging over reconstructions is denoted by $\langle\ \rangle$, and $\langle\ \rangle_\varphi$ corresponds to the averaging over azimuth angle. Alternatively, one can use the phase transfer function (Marchesini et al., 2005), which does not require the knowledge of the ideal scattering amplitude:

$$PTF(q) = \left\langle \left| \left\langle \frac{G(\mathbf{q})}{|G(\mathbf{q})|} \right\rangle \right| \right\rangle_\varphi$$

(18b)

Then the resolution can be evaluated from the TF scattering wave vector cut off. The plots of the TF, corresponding to different data acquisition times, and therefore different signal-to-noise ratios (SNR), are shown in Fig. 6.. The top row of images in Fig. 5 was used to obtain the curves in Fig. 6. Averaging of the reconstructed objects using phase information produces similar curves. The thick line (1) corresponds to the ideal diffraction intensity, being the input for the HIO procedure. It reflects the effects of imperfect phasing by the iterative algorithm itself, and exhibits a flat plateau at lower scattering vectors with a rather abrupt cut off, characteristic for a coherent imaging system. Other curves demonstrate the TF response to the introduction of shot noise. We determined the resolution limit for a given exposure time from the width $q_{1/2}$ of the corresponding TF at half maximum (TF = 0.5) as $d = 2\pi/q_{1/2}$. The results are shown in Fig. 7 in the form of a plot of data acquisition time as a function of resolution, for images averaged in reciprocal (solid squares) and real (open circles) space. Both sets of data points follow the power law at high resolution, but experience a sharp decrease at about 30 Å. This is especially obvious for the images averaged in real space, where apparent resolution becomes virtually independent of exposure. This effect is related to the specific features of the object structure, dominated by the well-defined rings with average periodicity of 40 Å. Therefore, at lower exposure time only the strong scattering due to these rings would be reliably detectable, even at very short counting times. In this case, the phasing algorithm does not properly retrieve the phases of the scattering amplitude, as demonstrated in Fig. 5 by the blurring of the images averaged in reciprocal space. But it still produces the distinctive strips, which do not vanish upon real space averaging. For the linear fit of the data we used only the five points giving the highest (best) resolution, where the required time obeys the power scaling with $d$. For the images, averaged in reciprocal space, the exponent of the power law is -3.7 (-3.6 for the real space averaging), in good agreement with Eq. (10).

For a quantitative comparison with the analytical results of section 2, we assume an empirical protein composition of $H_{50}C_{30}N_9O_{10}S_1$ and density 1.35 gm/cm$^3$, which gives an average electron density of 434 nm$^{-3}$. We also take into account that Eq. (2) is derived for critical sampling, and rescale it to the sampling ratio of the 256×256 grid according to $t \propto s^2$. The resolution predicted by Eq. (10) at a given time (dash dot line) is better by a factor of 1.5 than that derived from simulated images. This discrepancy may be due partially to the additional effect of shot noise on the phasing algorithm stability, and to the arbitrarily setting of the number of counts required for the data statistical accuracy, that appears to be too low, in the analytical solution. The possible reason is that the variation in scattered intensity rather than absolute count rate must be accurately measured, which would require a better SNR. Considering the count rate $P$ in the pixel at resolution limit as a free parameter in our analytic model, the resolution, determined from the TF calculation and shown in the Fig. 7 by solid squares, can be fitted by Eq. (10) with $P = 25$ counts/pixel. That is higher than the count rate expected from the Rose criterion. Fig. 7 also shows the required time dependencies on resolution described by Eq. (1) due to Howells et al. (2005) (dash line) and Eq. (2) by Shen et al. (2004) (dot line). These curves give too optimistic results for the expected resolution as compared to the explicit TF calculation. The resolution definition via TF is still uncertain due to its complicated shape and absence of a sharp cut off, emphasizing that for phase-contrast imaging, resolution cannot be specified by a single parameter, and depends on the sample itself. In Fig. 5 we observe that details of the shape envelope distort at counting times less than 100 s. Collecting data at the critical sampling ratio would reduce the required exposure by a factor of 12.6. However, we found that decreasing the sampling ratio reduces the stability of the HIO algorithm convergence to a valid solution. The addition of more constraints to the phasing algorithm due to an a-priori information may be available, especially convex ones such as the widely used histogram constraint. That may allow reduced oversampling.

Note that all calculations have been done for one sample in the beam. Application of a "shower-head" multiple nozzle aerojet array, which is currently under development, has the potential to increase the number of molecules simultaneously present in the beam to about 100. This would substantially reduce the time required for diffraction measurement.

In summary, when full account is taken of Poisson noise and the performance of the phasing algorithm, we find using Eq. (10) that the exposure time for Serial Crystallography is given by

$$\Delta t = \frac{1.3 \times 10^9 \; s^2}{M I_0 d^4 \lambda^2}$$

(19)

where we use nm and second units, and the scaling constant is derived from the Fig. 7. The severe dependence on the poorly defined resolution $d$ is noted. ($d$ is poorly defined because it depends on the structure of the sample. Our resolution definition using MTF=0.5 is highly conservative). This power law has serious implications for all attempts at coherent imaging with X-rays. Table 1 shows the estimates of the expected counting times at the planned ALS, APS and ERL X-ray beamlines from Eq. (19), which demonstrate a severe punishment in terms of the required exposure time for a very small resolution improvement. We note that, under the dose fractionation theorem of Hegerl and Hoppe (1976), these times are increased by a factor of 40 for 3D image data collection.

**5. Summary.**

The simple way to estimate the diffraction experiment counting time required for a given resolution is to calculate the number of photons scattered at the angle corresponding to this resolution, and to set this number to a fixed value, which would provide the statistically accurate measurement. We performed this calculation analytically for a globular uniform object and numerically by simulating the diffraction pattern for the chaperonin GroEL-GroES protein complex. This approach gives the lower limit of the required exposure. For a more elaborate evaluation, which also accounts for the convergence stability of the phase retrieval algorithm and its effect on resolution, we have used the HIO procedure to reconstruct charge density maps in real space from simulated diffraction patterns with different noise levels. Visual examination of the reconstructed images shows that at the projected ERL X-ray beam source even the short exposure of 10-100 s can produce valuable information on the bio-complex envelope shape.

Using the transfer function spatial frequency cut off as a quantitative measure of resolution, we determined the functional dependence of the exposure time on required resolution. It scales as the inverse forth power of *d*. However, the prefactor obtained by fitting to the calculated resolution is higher than that expected from the Rose criterion. Using the count rate, required for a statistically accurate measurement, as a free parameter, we get the exact agreement with the analytical solution.

The times predicted by the simple analytical models given here and by Henke and DuMond (1955), Howells et al. (2005), and Shen et al. (2004) can be up to two orders of magnitude shorter than those following from the TF calculation, since they do not include the effects of the phasing algorithm on resolution. These three analytical model treatments may be distinguished as follows.

1. In Henke and DuMond (1955) and Howells et al. (2005), a coherent sum of scattering from one voxel (resolution element) inside the sample is used. The result depends on which voxel is chosen.

2. In Shen et al. (2004) treatment, an incoherent sum over all voxels is used at the maximum (resolution limiting) scattering angle. Interference between waves scattered by different voxels is ignored by averaging, and the result depends on molecular size. We note a $d^{-3}$ scaling of exposure time in this approach.

3. In our treatment, a coherent sum over all voxels is used at the maximum (resolution limiting) scattering angle. The result again depends on the size of the molecule.

The reported results have important implications for the design of droplet beam systems for serial crystallography, suggesting that the use of multiple nozzles will be essential for third-generation synchrotrons, but not for fourth generation machines.

Possibilities for decreasing the exposure time required to achieve a desired resolution include use of lower X-ray energy, optimization of coherence conditions, increasing the number *M* of proteins present in the X-ray beam at any instant, use of a more efficient phasing algorithm (Marchesini, 2007) and use of additional constraints in the phasing algorithm, such as the histogram constraint (which drives the density map toward the known grey-level histogram for protein density maps), allowing smaller oversampling ratio *s*. Additional a-priori information may also be available, such as bond-lengths and sequence. The method of molecular replacement may also be useful, and has now succeeded in solving a protein structure from powder diffraction data (Von Dreele et al., 2000). Taken together, these improvements would reduce the required exposure at the ERL down to a value of a few tens of seconds.

### 6. Acknowledgements.

This work was supported by the ARO grant DAAD190010500 and NSF award IDBR 0555845.

**Table 1** Exposure time (s) required to achieve a given resolution at different X-ray beamlines with parameters, discussed in the text, calculated from Eq. (18) for $s = 2^{1/2}$ and $M = 10$. Reducing the spatial coherence at the ALS to match the molecular size would decrease the required exposure time by several orders of magnitude.

|  | $d = $ 0.7 nm | $d = $ 1 nm | $d = $ 2 nm |
|---|---|---|---|
| ALS | $4.2 \times 10^3$ | $1.0 \times 10^3$ | 63 |
| APS | $1.1 \times 10^4$ | $2.7 \times 10^3$ | 171 |
| ERL | 150 | 36 | 2.3 |

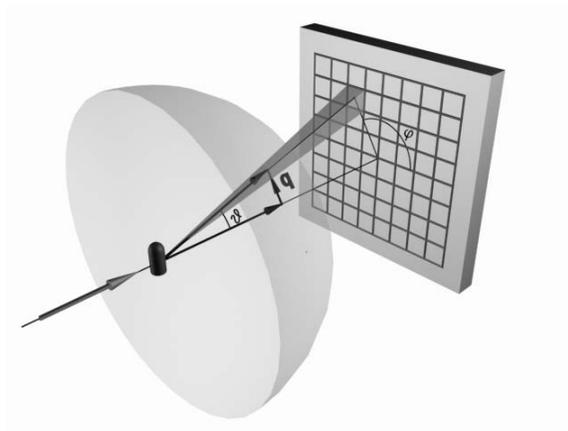

**Figure 1** Scattering geometry for simulation of diffraction pattern.

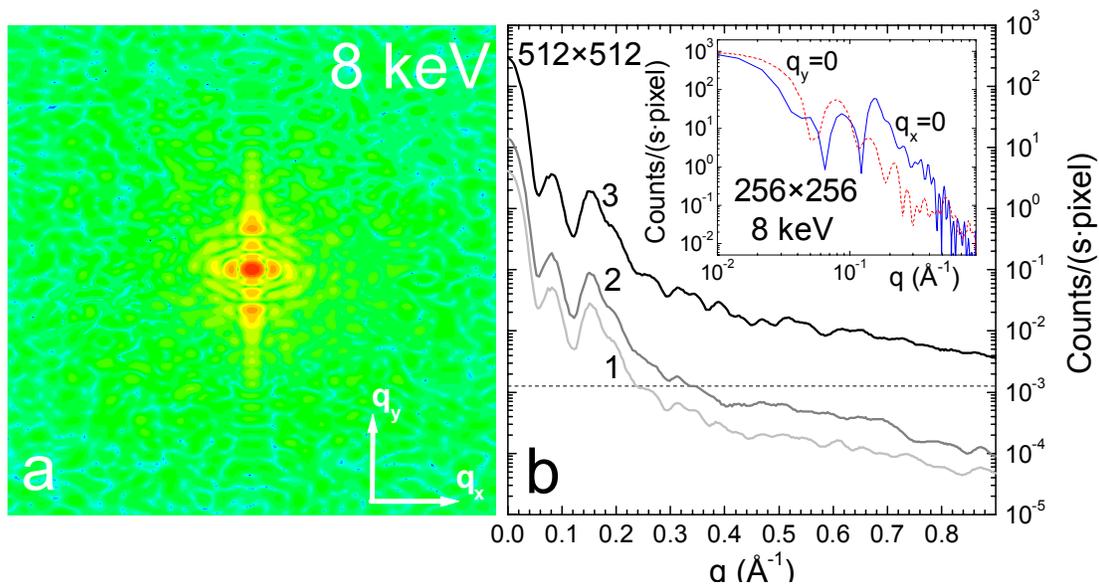

**Figure 2** (a) Diffraction pattern for GroEL complex at X-ray energy of 8 keV. (b) Scattered intensity per pixel after angular averaging at: (1) 5.4 keV and $1.8\times10^6$ photons s$^{-1}$ nm$^{-2}$ (APS); (2) 3.0 keV and $1.5\times10^6$ photons s$^{-1}$ nm$^{-2}$ (ALS); (3) 8.0 keV and $3\times10^8$ photons s$^{-1}$ nm$^{-2}$ (ERL). Inset shows the scattered counts per pixel for the incident flux (3) on the 256×256 grid, cut through the planes $q_x = 0$ (solid line) and $q_y = 0$ (dash line), indicated in (a).

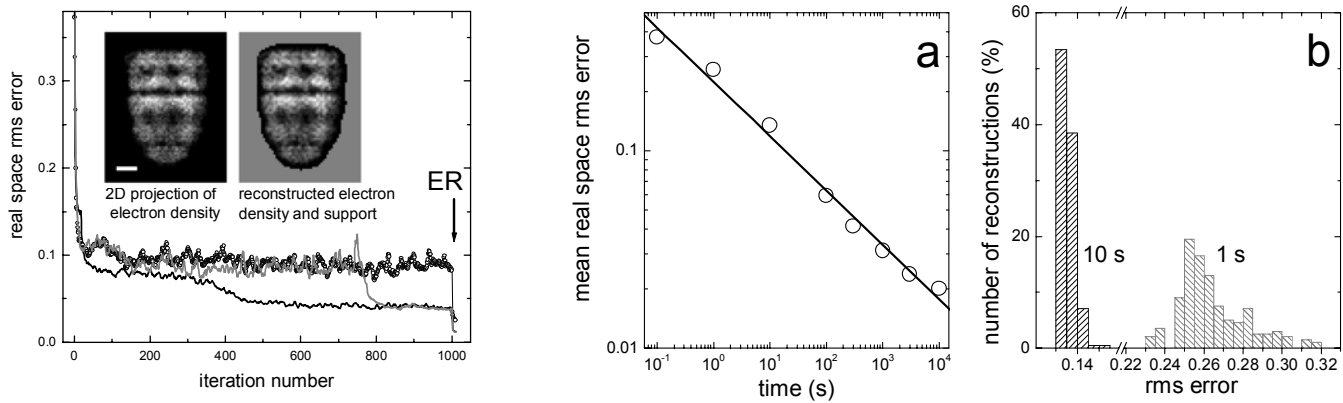

**Figure 3** Real space error for the HIO algorithm as a function of iteration cycle number. Solid lines show the error behavior for two successful runs, while circles correspond to the reconstruction, which did not converge to the solution. Arrow indicates the step where the ER algorithm was applied. Inset: the charge density projection of the protein complex used to calculate the diffraction pattern (left panel) and averaged reconstruction (right panel). Area outside the support is filled with gray color. The bar length is 35 Å (10 pixels).

**Figure 4** (a) Reconstruction rms error in real space, averaged over many independent reconstructions, as a function of counting time. Solid line shows the best fit to the data points by a power law. (b) Distribution of the rms error in two sets of independent reconstructions for the counting time of 10 s and 1 s.

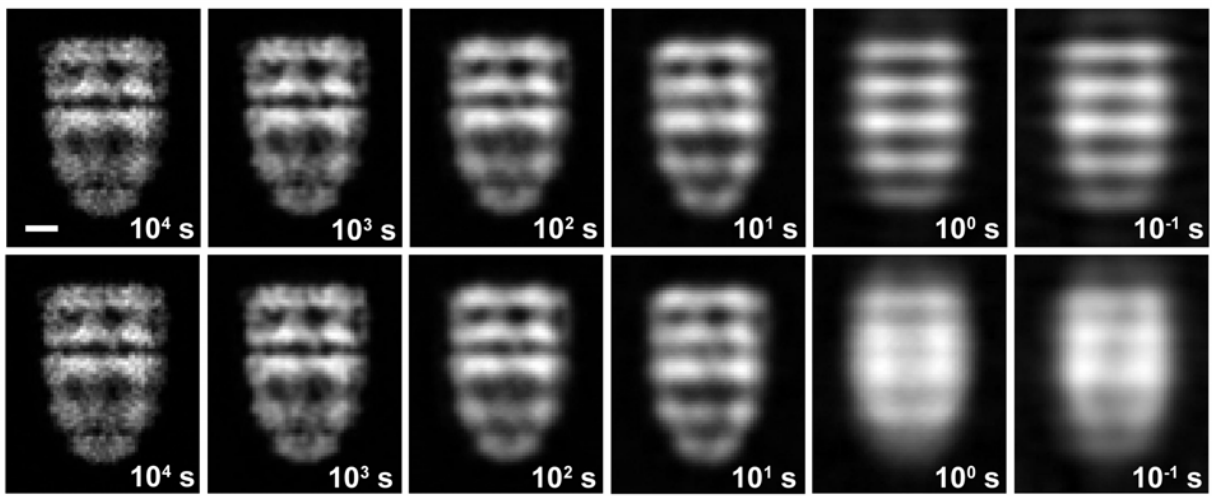

**Figure 5** The series of averaged reconstructed charge densities for the indicated exposure times. Before the averaging the images are aligned by translation in real space (top row) or by phase adjustment in reciprocal space (bottom row). The incident flux is $3\times10^8$ photons nm$^{-2}$ s$^{-1}$ and X-ray energy is 8 keV.

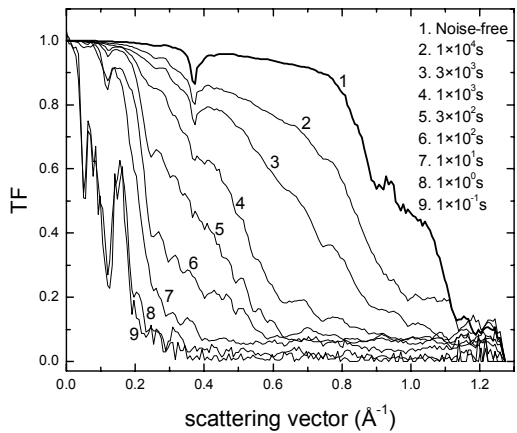

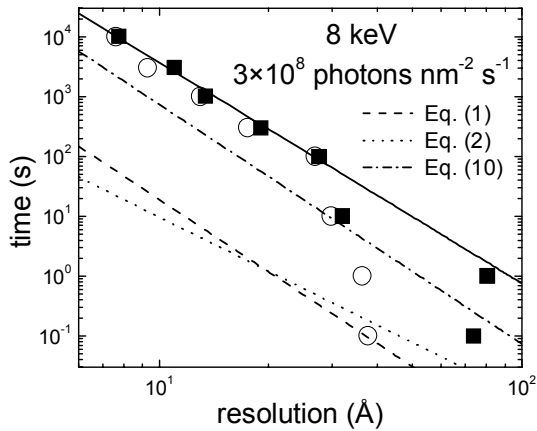

**Figure 6** HIO transfer function for different signal-to-noise ratios (exposure times).

**Figure 7** Exposure time requirement for a given resolution, deduced from the TF width for the HIO output, averaged using phase adjustment in reciprocal space (solid squares) and cross-correlation in real space (open circles). Solid line is the least-squares linear fit to the former data set for resolution higher than 30 Å. Other lines are given for comparison with simple analytical predictions from Howells et al. (2005) (dash line), Shen et al. (2004) (dot line) and this paper (dash-dot line) with parameter $P = 5$.